\documentstyle[12pt]{article}
\begin{document}
\baselineskip 15pt

\title{\bf Can the Decoherent Histories Description of Reality be
Considered Satisfactory?}
\author{Angelo Bassi\footnote{e-mail: bassi@ts.infn.it}\\ {\small
Department of Theoretical
 Physics of the University of Trieste,}\\ and \\
\\ GianCarlo Ghirardi\footnote{e-mail: ghirardi@ts.infn.it}\\ {\small
Department of
Theoretical Physics of the University of Trieste, and}\\ {\small the Abdus
Salam International
 Centre for Theoretical Physics, Trieste, Italy.}}

\date{}

\maketitle 

\def\tu{{t_{1}}}
\def\td{{t_{2}}}
\def\tt{{t_{3}}}
\def\tn{{t_{n}}}
\def\tm{{t_{m}}}
\def\pau{{P^{\alpha_{1}}_{1}}}
\def\pbu{{P^{\beta_{1}}_{1}}}
\def\pad{{P^{\alpha_{2}}_{2}}}
\def\pbd{{P^{\beta_{2}}_{2}}}
\def\pan{{P^{\alpha_{n}}_{n}}}
\def\panmu{{P^{\alpha_{n-1}}_{n-1}}}
\def\pbn{{P^{\beta_{n}}_{n}}}
\def\pbnmu{{P^{\beta_{n-1}}_{n-1}}}
\def\pam{{P^{\alpha_{m}}_{m}}}
\def\pbm{{P^{\beta_{m}}_{m}}}
\def\pak{{P^{\alpha_{k}}_{k}}}
\def\sux{{\sigma^{1}_{x}}}
\def\suy{{\sigma^{1}_{y}}}
\def\suz{{\sigma^{1}_{z}}}
\def\sdx{{\sigma^{2}_{x}}}
\def\sdy{{\sigma^{2}_{y}}}
\def\sdz{{\sigma^{2}_{z}}}
\def\uxp{{|1x^{+}\rangle}}
\def\uxm{{|1x^{-}\rangle}}
\def\dxp{{|2x^{+}\rangle}}
\def\dxm{{|2x^{-}\rangle}}
\def\uyp{{|1y^{+}\rangle}}
\def\uym{{|1y^{-}\rangle}}
\def\dyp{{|2y^{+}\rangle}}
\def\dym{{|2y^{-}\rangle}}
\def\uzp{{|1z^{+}\rangle}}
\def\uzm{{|1z^{-}\rangle}}
\def\dzp{{|2z^{+}\rangle}}
\def\dzm{{|2z^{-}\rangle}}
\def\puxppdxp{{P_{{1x}^{+}{2x}^{+}}}}
\def\puxppdxm{{P_{{1x}^{+}{2x}^{-}}}}
\def\puxmpdxp{{P_{{1x}^{-}{2x}^{+}}}}
\def\puxmpdxm{{P_{{1x}^{-}{2x}^{-}}}}
\def\puyppdyp{{P_{{1y}^{+}{2y}^{+}}}}
\def\puyppdym{{P_{{1y}^{+}{2y}^{-}}}}
\def\puympdyp{{P_{{1y}^{-}{2y}^{+}}}}
\def\puympdym{{P_{{1y}^{-}{2y}^{-}}}}
\def\puxppdyp{{P_{{1x}^{+}{2y}^{+}}}}
\def\puxppdym{{P_{{1x}^{+}{2y}^{-}}}}
\def\puxmpdyp{{P_{{1x}^{-}{2y}^{+}}}}
\def\puxmpdym{{P_{{1x}^{-}{2y}^{-}}}}
\def\puyppdxp{{P_{{1y}^{+}{2x}^{+}}}}
\def\puyppdxm{{P_{{1y}^{+}{2x}^{-}}}}
\def\puympdxp{{P_{{1y}^{-}{2x}^{+}}}}
\def\puympdxm{{P_{{1y}^{-}{2x}^{-}}}}
\def\pxypyxpzzp{{P_{(xy)^{+}(yx)^{+}(zz)^{+}}}}
\def\pxymyxmzzp{{P_{(xy)^{-}(yx)^{-}(zz)^{+}}}}
\def\pxymyxpzzm{{P_{(xy)^{-}(yx)^{+}(zz)^{-}}}}
\def\pxypyxmzzm{{P_{(xy)^{+}(yx)^{-}(zz)^{-}}}}
\def\pxxpyymzzp{{P_{(xx)^{+}(yy)^{-}(zz)^{+}}}}
\def\pxxmyypzzp{{P_{(xx)^{-}(yy)^{+}(zz)^{+}}}}
\def\pxxpyypzzm{{P_{(xx)^{+}(yy)^{+}(zz)^{-}}}}
\def\pxxmyymzzm{{P_{(xx)^{-}(yy)^{-}(zz)^{-}}}} \def\his{{\textsc{His}}}
\def\hisuxpdxp{{\his[1x^{+}2x^{+}]}}
\def\hisuxpdxm{{\his[1x^{+}2x^{-}]}}
\def\hisuxmdxp{{\his[1x^{-}2x^{+}]}}
\def\hisuxmdxm{{\his[1x^{-}2x^{-}]}}
\def\hisxxpp{{\his[(xx)^{+}]}}
\def\hisxxmm{{\his[(xx)^{-}]}}
\def\hisuypdyp{{\his[1y^{+}2y^{+}]}}
\def\hisuypdym{{\his[1y^{+}2y^{-}]}}
\def\hisuymdyp{{\his[1y^{-}2y^{+}]}}
\def\hisuymdym{{\his[1y^{-}2y^{-}]}}
\def\hisyypp{{\his[(yy)^{+}]}}
\def\hisyymm{{\his[(yy)^{-}]}}
\def\hisxypyxpzzp{{\his[(xy)^{+}(yx)^{+}(zz)^{+}]}}
\def\hisxymyxmzzp{{\his[(xy)^{-}(yx)^{-}(zz)^{+}]}}
\def\hisxymyxpzzm{{\his[(xy)^{-}(yx)^{+}(zz)^{-}]}}
\def\hisxypyxmzzm{{\his[(xy)^{+}(yx)^{-}(zz)^{-}]}}
\def\hisxxpyymzzp{{\his[(xx)^{+}(yy)^{-}(zz)^{+}]}}
\def\hisxxmyypzzp{{\his[(xx)^{-}(yy)^{+}(zz)^{+}]}}
\def\hisxxpyypzzm{{\his[(xx)^{+}(yy)^{+}(zz)^{-}]}}
\def\hisxxmyymzzm{{\his[(xx)^{-}(yy)^{-}(zz)^{-}]}}
\def\hisxypp{{\his[(xy)^{+}]}}
\def\hisxymm{{\his[(xy)^{-}]}}
\def\hisyxpp{{\his[(yx)^{+}]}}
\def\hisyxmm{{\his[(yx)^{-}]}}
\def\hiszzpp{{\his[(zz)^{+}]}}
\def\hiszzmm{{\his[(zz)^{-}]}}
\def\hisuxpdyp{{\his[1x^{+}2y^{+}]}}
\def\hisuxpdym{{\his[1x^{+}2y^{-}]}}
\def\hisuxmdyp{{\his[1x^{-}2y^{+}]}}
\def\hisuxmdym{{\his[1x^{-}2y^{-}]}}
\def\hisuypdxp{{\his[1y^{+}2x^{+}]}}
\def\hisuypdxm{{\his[1y^{+}2x^{-}]}}
\def\hisuymdxp{{\his[1y^{-}2x^{+}]}}
\def\hisuymdxm{{\his[1y^{-}2x^{-}]}}
\def\hisu{{\his^{(1)}}}
\def\hisd{{\his^{(2)}}}
\def\hist{{\his^{(3)}}}
\def\fam{{\textsc{Fam}}}
\def\hisa{{{\textsc{His}}^{(\alpha)}}}

\begin{abstract}

We discuss some features of the Decoherent Histories approach. We consider
four assumptions, 
the first three being in our opinion necessary for a sound interpretation
of the theory, while the fourth one is accepted by the supporters of the DH
approach,
and we prove that they lead to a logical contradiction. We discuss the
consequences of relaxing any one of them.  

\end{abstract}

\section{Introduction.}

The Decoherent Histories (DH) approach of Griffiths \cite{gri1, gri2},
Omn\`es \cite{omn1,
omn2} and Gell--Mann and Hartle \cite{gel1, gel2, gel3} has attracted in
recent years a lot of
attention since it seemed to yield a solution to the
conceptual and interpretative problems of standard quantum mechanics (SQM)
without requiring
relevant changes of the formalism. This feature is not shared by other
attempts to work
out \cite{gold} {\it a quantum theory without observers}
like hidden variable theories \cite{bohm,gdz} which need additional
parameters besides
(or in place of) the wave function to characterize the state of an
individual physical system,
or by the dynamical reduction models \cite{GRW,P,GPR,GGB} which accept that
 Schr\"{o}dinger's
equation must be modified.

	The general structure of the theory can be summarized as follows:
let $S$ be a physical system
which at the initial time $t_{0}$ is associated to the statistical operator
$W(t_{0}) = W$,
and let $U(t,t')$ be the unitary operator describing its evolution. One
then chooses $n$
arbitrary times $t_{1} < t_{2} < \ldots < t_{n}$, and for each of them (let
us say $t_{m}$)
one considers an exhaustive set $\{\pam\}$ of mutually exclusive projection
operators:
\begin{equation}
	\sum_{\alpha_{m}}\pam = 1, \qquad\quad \pam\pbm =
	\delta_{\alpha_{m},\beta_{m}}\pam.
\end{equation}
One also considers the following projection operators:
\begin{equation} \label{e1}
	Q^{\alpha_{m}}_{m} = \sum_{\alpha_{m}} \pi^{\alpha_{m}}_{m} \pam,
\end{equation}
where $\pi^{\alpha_{m}}_{m}$ take the values 0 or 1.

One history is then defined by the sequence of times $\tu, \td,\ldots, \tn$
and a corresponding sequence of projection operators, each of them taken
from the family $\{Q^{\alpha_{m}}_{m}\}$ $(m = 1,2,...,n)$, defined as in
(\ref{e1}): 
\begin{equation} \label{W}
	\his^{(\alpha)} = \{(Q^{\alpha_{1}}_{1},\tu), 
	(Q^{\alpha_{2}}_{2},\td), \ldots,
	(Q^{\alpha_{n}}_{n},\tn)\}. \label{n1}
\end{equation}
When the operators appearing in eq. (\ref{W}) belong to the exhaustive set 
$\{\pam\}$, $\his^{(\alpha)}$ is said to be {\it maximally fine--grained}.
Consideration of the operators  $\{Q^{\alpha_{m}}_{m}\}$ corresponds to
taking into account {\it coarse--grained histories}. A family of histories
is a set whose elements are all hystories having the form (\ref{n1}), i.e.
all maximally fine-grained histories and their coarse-grainings. For a given
family one then considers what is usually denoted as the {\it decoherence
functional}\footnote{Note that here we consider only maximally fine-grained
histories.}
\cite{gel1}:
\begin{eqnarray} \label{n2}
	D(\alpha,\beta) & = & Tr[\pan U(t_{n},
	t_{n-1})\panmu U(t_{n-1},t_{n-2})\ldots 	U(t_{1},t_{0})W
\nonumber \\
	& & U^{\dagger}(t_{1},t_{0})\ldots U^{\dagger}(t_{n-1},t_{n-2})\pbnmu
	U^{\dagger}(t_{n}, t_{n-1})\pbn],
\end{eqnarray}
in which the 
projection operators $\pau,\ldots,\pan$ characterize the history
$\his^{(\alpha)}$
and the projection operators $\pbu,\ldots,\pbn$ another history
$\his^{(\beta)}$.
 A family of histories is said to be decoherent if and only if:
\begin{equation} \label{n3}
	D(\alpha,\beta) = \delta_{\alpha,\beta}D(\alpha,\alpha), \end{equation}
i.e. iff the decoherence functional vanishes when the two maximally
fine-grained histories
$\his^{(\alpha)}$ e $\his^{(\beta)}$ do not coincide. When they coincide, the
expressions $D(\alpha,\alpha)$  are assumed to define a probability distribution
over the maximally fined grained histories of the decoherent family. In
this case one can also consider the expressions corresponding to
$D(\alpha,\alpha)$ in which the coarse grained operators replace the fine
grained ones and assume that they give the probability of
the coarse-grained histories.  We will say that a history is decoherent if
it belongs to at least one decoherent family. 

As it should be
 clear from this presentation, the theory, at its fundamental level, does
not attach any
particular role either to measurement processes (even though it is
perfectly legitimate
to build up histories describing the unfolding with time of such processes
and the occurrence
of their outcomes), or to wave packet reduction, and represents
an attempt to get rid of all those features which make fundamentally
unsatisfactory the
 Copenhagen interpretation of SQM.

In this letter, we will discuss some interpretational issues of the DH 
approach, with special
regard to ``scientific
 realism''. In the literature analyses of a similar kind have been
presented
\cite{de1, de2, dk, k} by some authors and the supporters of
the DH approach
 have answered \cite{gri2, om1, gh1} to the remarks of the above quoted
papers. We will add
new arguments identifying precise problems that such an approach has to face:
in particular, in the next section we will put forward four assumptions,
the first three being,
 in our opinion,  necessary for a realistic interpretation of the
theory, while the fourth one is generally accepted by the supporters of the
DH approach. We will then prove, in section 3, that such assumptions lead
to a logical contradiction and thus they cannot hold simultaneously. In the
final
sections, we will analyze the consequences of relaxing any one of them.
Though we are aware that the
 supporters of the DH approach will not subscribe all our
assumptions, our
 argument helps to undertstand what can be done and, more important, what
cannot be done
 within the DH approach.

\section{Four assumptions.}

	Let us list explicitly our
four assumptions, and discuss their conceptual status. For more details we
refer the reader
to \cite{bg}. \\

{\bf a) Decoherent Families and Boolean Algebras.} Among the proponents and the
supporters of the Decoherent Histories, Omn\`es \cite{omn1}, and subsequently
 Griffiths \cite{gri1}, have suggested to equip any family of decoherent
histories with an
algebraic Boolean structure.
For simplicity (and also since in what follows we will always make reference to
families of this type) let us consider a family of histories characterized
by only one
time $t$, and, accordingly, by a unique exhaustive and exclusive
set of projection operators $\{P_{\alpha}\}$ and their grainings
$\{Q^{\alpha_{m}}_{m}\}$:
 \[ \his^{(\alpha)} =
\{(Q_{\alpha}, t)\}. \]
In such a case, the logical connectives, the conjunction and the
disjunction of two histories
and the negation of one history, are defined in the following way:
\begin{eqnarray*}
\his^{(\alpha)}\wedge\his^{(\beta)} =
\{(Q_{\alpha}\wedge Q_{\beta}, t)\}
& \qquad & Q_{\alpha}\wedge Q_{\beta} =
Q_{\alpha}Q_{\beta}, \\
\his^{(\alpha)}\vee\his^{(\beta)} =
\{(Q_{\alpha}\vee Q_{\beta}, t)\}
& \qquad & Q_{\alpha}\vee Q_{\beta} =
Q_{\alpha} + Q_{\beta} -
Q_{\alpha}Q_{\beta},
\\
\neg\his^{(\alpha)}\quad =\quad\,\,\,\,\, \{(Q_{\alpha}^{\perp}, t)\} & &
Q_{\alpha}^{\perp} = 1 - Q_{\alpha}.
\end{eqnarray*}
We stress that
the fact that any family can be equipped with a Boolean structure plays an
essential role
 within the theory since it guarantees that one can use the rules of
classical logic to deal
 with the histories belonging to a {\it single} decoherent family. This in
turn implies that
 the same rules can be used to argue about the physical properties
described by the histories,
 avoiding in this way
all difficulties characterizing quantum logics. Since
there is a general
 agreement about it, we will not discuss this feature any further.
\\

{\bf b) Decoherent Histories and truth values.} Let us restrict ourself to
a specific family of
histories.
As already stated, it is
one of the basic assumptions of the theory that, if the family satisfies
the decoherence
 conditions, the diagonal elements $D(\alpha, \alpha)$ of the decoherence
functional acquire
 the status of a {\it probability} distribution over the histories of the
decoherent family.
 In connection with such an assumption one is naturally led to raise the
question: {\it probability of what}? Of course, not probability of {\it
finding} the system
 having the properties described by the history $\hisa$ if a measurement is
performed,
otherwise the theory would not represent an improvement of the Copenhagen
interpretation
 of Quantum Mechanics. The only possible answer, in order to have a sound
theory, is that the {\it probabilities} refer to {\it objective properties
of the physical system}, like classical probabilities. Only
in this way one can hope to construct a realistic interpretation of Quantum
Mechanics, as
often advocated by the supportes of the DH approach.

To clarify this point, we can consider the probabilities of Classical
Statistical Mechanics. Such a
theory yields, in general, only a probability distribution over the sets of
subsets of the
phase space. Nevertheless, the theory allows to consider the physical
system one is interested
in as uniquely associated to a precise point in phase space at any given
chosen time\footnote{Actually, the theory specifies that the most 
complete characterization of the state of a physical system is just given by
the assignment of such a point in phase space. This is nothing but the
completeness requirement of the theory.}. 
This
association renders automatically true or false any statement concerning
the properties of the
 system\footnote{Note that the mentioned properties might also refer to a
certain graining,
such as ``the energy of a molecule lies within such and such an
interval...''}; actually it
is precisely this feature of Classical Statistical Mechanics that makes the
theory compatible
with a realistic attitude towards physical reality. For example, for a gas
we usually know
only its macroscopic thermodynamic properties like pressure, temperature, ...,
or the average value
of the microscopic ones; however, any statement concerning such
properties has a precise truth value (in general unknown to us) which is
uniquely determined
by the point representing the actual state of the system.

 On the other hand, in Standard Quantum Mechanics with the completeness
assumption, this is no more possible:
when a system is
in a
superposition of two states, one may not even think that it possesses one
of the properties
described by those two states. That is why the Copenhagen interpretation is
not a realistic
one; and that is why one needs the projection postulate in order to
actualize the quantum
potentialities through measurement processes.

Thus, in order to avoid giving up the request of realism, 
the probabilities of the DH approach must
be the analogous of classical probabilities: this means that to every
decoherent history it
must be possible to associate a precise truth
value 1 or 0, even though in general we do not know which one is the right
one, like in
 classical statistical mechanics.

Obviously, this assignement of truth values to histories can be expressed
formally by an
appropriate
homomorphism {\it h} from the histories of any {\it single} decoherent
family onto the
set $\{0,1\}$
which must satisfy the conditions making legitimate to resort to classical
reasoning
when dealing with such histories:
\begin{eqnarray*}
	h[\his^{(\alpha)}\wedge\his^{(\beta)}] & = &
	h[\his^{(\alpha)}]\wedge h[\his^{(\beta)}],
	\\ h[\his^{(\alpha)}\vee\his^{(\beta)}] & = & 	h[\his^{(\alpha)}]\vee
	h[\his^{(\beta)}], \\
	h[\neg\his^{(\alpha)}] & = &
 h[\his^{(\alpha)}]^{\perp} = 1 - h[\his^{(\alpha)}].
\end{eqnarray*}
In simpler terms, the homomorphism must preserve the logical operations of
conjunction,
 disjunction and negation. For instance if a history is true the fact that the
correspondence $h$ be an homomorphism satisfying the above relations
implies that its negation is false;
if one history is true and a second history is false, then their conjunction is
false, while
their disjunction is true. As already remarked, the homomorphic nature of
$h$ guarantees that
classical logic can be used within a {\it single} family of decoherent
histories and that
the truth values associated to the elements (i.e. the histories) of the
boolean algebras (i.e.
the decoherent families) obey classical rules. \\

{\bf c) Histories belonging to different families.} According to the
proponents of the DH
 approach, it is one of the firm points of the theory that one cannot
compare different
 histories belonging to decoherent families which are incompatible among
themselves, i.e.
such that there does not exist a decoherent family which can accomodate all
of them. Thus
 any conclusion one derives from such histories is neither true nor false;
it is simply devoid
of any meaning. One cannot however avoid
raising the following question: when the {\it same} history belongs to {\it
different}
decoherent families (which are generally incompatible), should one require
that its truth
 value be the same or should one allow it to change when he changes the
family? Or is this
 question meaningless, within the theory? We remember that we previously
said that any
decoherent history should have a truth value and, as such, should be
related to some
``element of reality'' when it is true. Then, if one accepts that the truth
value
of {\it the same history} depends on the decoherent family
to which it is considered to belong, one has to face an extremely
embarrassing situation:
if he looks at the history from the perspective of a given decoherent
family, then it may
 turn out (e.g.) to be true, i.e. to represent properties objectively
possessed by the
physical system at the times characterizing the history.
But, alternatively, if he considers the same history as belonging to a
second decoherent
family (different from the previous one) then it may turn out to be false,
i.e., it
identifies physical properties which {\it are not} possessed by the
physical system at
the considered times. {\it Nor can one say that the question we have posed
is not legitimate}
 (unless he denies from the beginning the very existence of truth values),
since when we are
talking about decoherent histories,  we are
talking about physical properties that systems possess or fail to possess,
and it is
important to know whether these properties are objective or depend in some
way upon our
choice of the family.
We believe that it is unavoidable to assume that the truth value of a
single history cannot
 depend from the decoherent family one is considering. It seems to us
that this assumption is {\it fundamental} in order to have a {\it realistic}
interpretation of
 physical processes. Anyway, we will further comment on this crucial point
in the final Sections.\\

{\bf d) How many families of decoherent histories can be considered?} One
of the
main difficulties that the theory had to face since its appearence is the
following:
 are {\it all} families of decoherent histories equally legitimate to
describe objective
 properties of physical systems or should one
introduce some criterion limiting the number of acceptable families to few,
or even to only
 one of them? The fundamental reasons for which this problem has to be
faced are the
following. First, the very existence of incompatible decoherent families
gives rise to
various difficulties of interpretation; as already remarked
different histories belonging to
 incompatible families, when considered separately can be assumed to
describe correctly
the properties of a physical system, while it is forbidden to consider them
together.
 This feature of the theory seems absolutely natural to the supporters of
the Decoherent
Histories, but it is a source of worries for
the rest of the scientific community. Secondly, there are many families
(actually the
majority of them) which, in spite of the fact that they satisfy the
decoherence condition,
cannot be endoved by any direct physical meaning:
how can then one consider them as representing objective properties of
physical systems?
In spite of these difficulties, some supporters \cite{gri2} insist in
claiming that there are no privileged families. Accordingly, we take (for
the moment)
the same point of view and we assume that any decoherent family has to be
taken into account.
 Actually, in the proof of our theorem, we will limit our considerations to
very few and quite
reasonable families, and we will by no means need to resort to the
consideration of exotic
histories to derive our conclusions. \\

In the next section we prove that the Decoherent Histories approach of
quantum mechanics, to avoid logical inconsistencies, requires to give up at
least one of
 the previous four assumptions. We
will not exhibit
the  general derivation of such a conclusion (which has been given in
\cite{bg}), but we
will prove our theorem with reference to a quite simple example which is
sufficient to make
 clear the crucial lines of our reasoning.

\section{An explicit example proving that the Decoherent Histories approach
is incompatible
with the four previous assumptions.}

Let us focus our attention on a quite simple physical system, i.e. two spin
$1/2$
particles. We take into account only the spin degrees of freedom and we
suppose that the
Hamiltonian does not involve the spin variables (so that one can consider
it as identically
equal to zero --- the quantum state of the system does not change with
time). Let us consider
the spin operators $\sux$, $\suy$, $\suz$ (in units of $\hbar/2$) for
particle 1,
and $\sdx$, $\sdy$, $\sdz$ for particle 2.

We take now into account the following table of nine spin operators for the
composite system:

\begin{center}
\begin{tabular}{ccc}
	& & \\
	\quad $\sux$ \quad
& \quad $\sdx$ \quad & \quad $\sux\sdx$ \quad \\ 	& & \\
	& & \\
	\quad $\sdy$ \quad & \quad $\suy$ \quad & \quad $\suy\sdy$ \quad \\
	& & \\
	& & \\
	\quad $\sux\sdy$ \quad & \quad $\suy\sdx$ 	\quad & \quad$
\suz\sdz$\quad \\
	& & \\
\end{tabular}
\end{center}

This set of operators has been first considered by Peres \cite{Per} and
Mermin \cite{Mer},
to investigate the unavoidable contextuality of any deterministic hidden
variable theory.
 Their argument is quite straightforward: if one assumes that the
specification of the hidden
variables determine {\it per se} which one of the two possible values (+1
and -1) these
operators ``possess'', one gets a contradiction. In fact, since the
product of the three {\it commuting} (and thus {\it compatible}) operators
of each line
and of the first two columns is the identity operator (which must
obviously assume the value 1 for any choice of the hidden variables) while
the product of
the three {\it commuting} operators of the last column equals minus the
identity operator,
no acceptable assignement of values (+1 and -1) to the nine operators can
be made. The way
out from this difficulty is also well known: one has to accept the
contestual nature of
 possessed properties, meaning that the truth value of (e.g.)
the statement ``this observable has the value +1'' is not uniquely
determined by the complete
 specification of the system under consideration but it depends on the
overall context. In
the case under consideration this means that the truth value of the considered
statement might (and actually for at least one of them must) depend on the
fact that
the considered observable is measured together with the others {\it
compatible} observables
appearing in the same line, or together with the
others {\it compatible} observables of the column to which it belongs. This
fact
is considered as puzzling by some people and absolutely natural by others
\cite {GDZErk}.
In any case, the way out does not lead to inconsistencies since some of the
operators
 appearing
in the considered line and column do not commute among themselves. It is
therefore impossible
 to perform simultaneously the two sets of experiments. We would like to
stress the crucial
fact that the ambiguity about the truth values is here directly associated
to actual
physically different situations. In the words of the authors
of \cite {GDZErk} this fact {\it reflects little more than the rather
obvious observation
that the result of an experiment should depend upon how it is performed!}

We now work out an argument related to the one just mentioned, which
however has a completely
 different conceptual status since it deals with a theoretical scheme in
which there are no
 hidden variables; besides, we will always deal with projection operators
of a set of
 observables such that the resulting families are decoherent --- this makes
the proof
 more lengthly than in the previous case.

We consider six families of decoherent histories all of them being
one--time histories
 referring to the same time instant $t
>t_{0}$
($t_{0}$ being the initial time)
and to the same initial state described by a given statistical operator
(which we do not
 need to specify). Being one--time histories the corresponding families are
characterized
 by one exhaustive set $\{\pam\}$ of mutually exclusive projection
operators plus their coarse-grainings and they turn
out to be automatically decoherent. Let us characterize them in a precise
way: \\

$\bullet$ {\bf Family A.} The histories of this family make reference to
the properties of
 the observables $\sux$, $\sdx$ and $\sux\sdx$. Since such operators
commute with each other
one can characterize the maximally fine--grained histories of the family as
those associated
to the projection operators on their common eigenmanifolds .
Let us list the common eigenstates, the corresponding eigenvalues and the
associated
projection operators and histories:

a) The first eigenstate is:

\[ \uxp\otimes\dxp\quad \Longrightarrow\quad \left\{ \begin{array}{cl} 	+1
& \sux \\ +1 &
 \sdx \\ +1 & \sux\sdx \end{array} \right. \] the associated projection
operator is
 $\puxppdxp$ and the history corresponding to it will be denoted as
$\hisuxpdxp$ .

b) The second one is:

\[ \uxp\otimes\dxm\quad \Longrightarrow\quad \left\{ \begin{array}{cl} +1 &
\sux \\ -1 &
 \sdx \\ -1 & \sux\sdx\end{array} \right. \] whose associated projection
operator is
 $\puxppdxm$ and the corresponding history $\hisuxpdxm$.

c) The third eigenstate is:

\[ \uxm\otimes\dxp\quad \Longrightarrow\quad \left\{ \begin{array}{cl}
 -1 & \sux \\ +1 & \sdx \\ -1 & \sux\sdx\end{array} \right. \]
whose associated projection operator is $\puxmpdxp$ and the corresponding
history
 $\hisuxmdxp$.

d) Finally, the fourth common eigenstate is:

\[ \uxm\otimes\dxm\quad \Longrightarrow\quad \left\{ \begin{array}{cl} 	-1
& \sux \\ -1 & \sdx \\ +1 & \sux\sdx
	\end{array} \right. \]
whose
associated projection operator is $\puxmpdxm$ and the corresponding history
$\hisuxmdxm$.

Besides the four histories we have just listed it is useful, for our future
purposes,
to take into account the two following coarse--grained histories:

\begin{equation}
	\hisxxpp\quad =\quad\hisuxpdxp\vee\hisuxmdxm \label{eq1}
\end{equation}
\begin{equation}
	\hisxxmm\quad
=\quad\hisuxpdxm\vee\hisuxmdxp \label{eq2} \end{equation}
Obviously, the
first of these histories is associated to the projection operator
$\puxppdxp$ + $\puxmpdxm$.
Note that if this history is true, then the property possessed by the
system referring to the
 operator $\sux\sdx$ is the one corresponding to the eigenvalue +1, while,
if it is false, it
is the one corresponding to the eigenvalue -1. The second coarse--grained
history is associated
 to the projection operator
$\puxppdxm$ + $\puxmpdxp$, and it corresponds to the negation of the
history $\hisxxpp$. \\

$\bullet$ {\bf Family B.} It deals with properties related to the operators
$\suy$, $\sdy$ e $\suy\sdy$. The game is strictly analogous to the previous
one: the
basic histories
being $\hisuypdyp$ associated to the projection operator $\puyppdyp$;
$\hisuypdym$ associated
to the projection operator $\puyppdym$; $\hisuymdyp$ associated to the
projection
 operator $\puympdyp$, and, finally, $\hisuymdym$ associated to the
projection operator
 $\puympdym$. We will also deal with the two coarse--grained histories:
\begin{equation}
	\hisyypp\quad =\quad\hisuypdyp\vee\hisuymdym \label{eq3} \end{equation}
\begin{equation}
	\hisyymm\quad =\quad\hisuypdym\vee\hisuymdyp \label{eq4} \end{equation}
which are associated to the projection operator $\puyppdyp$ + $\puympdym$
and to the
eigenvalue $+1$ of the operator $\suy\sdy$; and to the projection operator
 $\puyppdym$ + $\puympdyp$, corresponding to the negation of the previous
history, respectively.
\\

$\bullet$ {\bf Family C.} The relevant commuting operators are $\sux\sdy$,
$\suy\sdx$
and $\suz\sdz$, their common eigenstates and the corresponding eigenvalues
are:

a) the first one is:

\[ \frac{1}{\sqrt{2}}[\uzp\otimes\dzp + i\uzm\otimes\dzm]\quad
\Longrightarrow\quad
\left\{ \begin{array}{cl}
	+1 & \sux\sdy \\ +1 & \sdy\sdx \\ +1 & \suz\sdz 	\end{array}
\right. \]
The associated projection operator is $\pxypyxpzzp$ and the
corresponding
history \\ $\hisxypyxpzzp$.

b) The second one is:
\[ \frac{1}{\sqrt{2}}[\uzp\otimes\dzp - i\uzm\otimes\dzm]\quad
\Longrightarrow\quad
\left\{ \begin{array}{cl}
	-1 & \sux\sdy \\ -1
& \sdy\sdx \\ +1 & \suz\sdz
	\end{array} \right. \]
The associated
projection operator is $\pxymyxmzzp$ and the corresponding history \\
$\hisxymyxmzzp$.

c) The third is:

\[ \frac{1}{\sqrt{2}}[\uzp\otimes\dzm + i\uzm\otimes\dzp]\quad
\Longrightarrow\quad
\left\{ \begin{array}{cl}
	-1 & \sux\sdy \\ +1
& \sdy\sdx \\ -1 & \suz\sdz
	\end{array} \right. \]
The associated
projection operator is $\pxymyxpzzm$ and the corresponding history \\
$\hisxymyxpzzm$.

d) Finally, the fourth one is:

\[ \frac{1}{\sqrt{2}}[\uzp\otimes\dzm - i\uzm\otimes\dzp]\quad
\Longrightarrow\quad
\left\{ \begin{array}{cl}
	+1 & \sux\sdy \\ -1
& \sdy\sdx \\ -1 & \suz\sdz
	\end{array} \right. \]
The associated
projection operator is $\pxypyxmzzm$ and the corresponding history \\
$\hisxypyxmzzm$.

We will also consider the following six coarse-grained histories:
\begin{equation}
	\hisxypp\quad =\quad\hisxypyxpzzp\vee\hisxypyxmzzm, \label{eq5}
\end{equation}
\begin{equation}
	\hisxymm\quad =\quad \hisxymyxmzzp\vee\hisxymyxpzzm, \label{eq6}
\end{equation}
\begin{equation}
	\hisyxpp\quad =\quad \hisxypyxpzzp\vee\hisxymyxpzzm, \label{eq7}
\end{equation}
\begin{equation}
	\hisyxmm\quad =\quad \hisxymyxmzzp\vee\hisxypyxmzzm, \label{eq8}
\end{equation}
\begin{equation}
	\hiszzpp\quad =\quad \hisxypyxpzzp\vee\hisxymyxmzzp, \label{eq9}
\end{equation}
\begin{equation}
	\hiszzmm\quad =\quad \hisxymyxpzzm\vee\hisxypyxmzzm. \label{eq10}
\end{equation}
According to
the above definition we have:
\begin{equation}
\hiszzpp\quad
	=\quad\{\hisxypp\wedge\hisyxpp\}\vee\{\hisxymm\wedge\hisyxmm\},
	\label{eq11}
\end{equation}
\begin{equation}
	\hiszzmm\quad
	=\quad\{\hisxymm\wedge\hisyxpp\}\vee\{\hisxypp\wedge\hisyxmm\},
	\label{eq12}
\end{equation}
and, obviously, the corresponding
relations hold for their images under
the homomorphisms. \\

$\bullet$ {\bf Family D.} It accomodates the operators $\sux$, $\sdy$ and
$\sux\sdy$.
The four maximally fine-grained histories are: $\hisuxpdyp$ whose
associated projection
 operator is $\puxppdyp$; $\hisuxpdym$ whose associated projection operator
is $\puxppdym$;
 $\hisuxmdyp$ whose
associated projection operator is $\puxmpdyp$; and finally history
$\hisuxmdym$ whose
associated projection operator is $\puxmpdym$. We will also deal with the
two following
coarse-grained histories:
\begin{equation}
	\hisxypp\quad =\quad \hisuxpdyp\vee\hisuxmdym, \label{eq13}
\end{equation}
\begin{equation}
	\hisxymm\quad =\quad \hisuxpdym\vee\hisuxmdyp. \label{eq14}
\end{equation}
As it is evident these histories are the same as those ((\ref{eq5}) and
(\ref{eq6}))
appearing in Family C.
In fact they are associated to the projection operators on the
eigenmanifolds of the
 operator $\sux\sdy$ corresponding to the eigenvalues
$+1$ and $-1$, respectively. According to assumption c), since these are
the same
histories, also their truth values will be the same. \\

$\bullet$ {\bf Family E.} It deals with the operators $\suy$, $\sdx$ and
$\suy\sdx$. The
four maximally fine-grained histories are:
$\hisuypdxp$,
whose associated projection operator is $\puyppdxp$; $\hisuypdxm$ whose
associated
 projection operator is
$\puyppdxm$; $\hisuymdxp$ whose
associated projection operator is $\puympdxp$; and finally the history
$\hisuymdxm$ whose
associated projection operator is $\puympdxm$. As usual we will also
consider two
 coarse--grained histories:
\begin{equation}
	\hisyxpp\quad =\quad \hisuypdxp\vee\hisuymdxm, \label{eq15}
\end{equation}
\begin{equation}
	\hisyxmm\quad =\quad \hisuypdxm\vee\hisuymdxp. \label{eq16}
\end{equation}
In this case these two histories coincide with the two coarse--grained
histories ((\ref{eq7})
 and (\ref{eq8})) belonging to Family C,
since they are identified by the same projection operators. Accordingly
the corresponding truth values must be the same. \\

$\bullet$ {\bf Family F.} This is the last family we will take into account
and it is
associated to the operators
$\sux\sdx$, $\suy\sdy$ and $\suz\sdz$. Once more the common eigenstates are:

\[ \frac{1}{\sqrt{2}}[\uzp\otimes\dzp + \uzm\otimes\dzm]\quad
\Longrightarrow\quad
\left\{ \begin{array}{cl} +1 & \sux\sdx \\ -1 & \sdy\sdy \\ +1 & \suz\sdz
	\end{array}
 \right. \]
whose associated projection operator is $\pxxpyymzzp$ and the corresponding
history
 $\hisxxpyymzzp$,
\[ \frac{1}{\sqrt{2}}[\uzp\otimes\dzp -
	\uzm\otimes\dzm]\quad \Longrightarrow\quad \left\{
\begin{array}{cl} 	-1
 & \sux\sdx \\ +1 & \sdy\sdy \\ +1 & \suz\sdz 	\end{array} \right. \]
whose associated projection operator is $\pxxmyypzzp$ and the corresponding
history
 $\hisxxmyypzzp$,
\[ \frac{1}{\sqrt{2}}[\uzp\otimes\dzm +
\uzm\otimes\dzp]\quad \Longrightarrow\quad \left\{ \begin{array}{cl} +1 &
\sux\sdx
 \\ +1 & \sdy\sdy \\ -1 & \suz\sdz \end{array} \right.\] whose associated
projection
 operator is $\pxxpyypzzm$ and the corresponding history $\hisxxpyypzzm$ ,

\[ \frac{1}{\sqrt{2}}[\uzp\otimes\dzm -
\uzm\otimes\dzp]\quad \Longrightarrow\quad \left\{ \begin{array}{cl} -1 &
\sux\sdx
 \\ -1 & \sdy\sdy \\ -1 & \suz\sdz\end{array} \right.\] whose associated
projection
operator is $\pxxmyymzzm$ and the corresponding history $\hisxxmyymzzm$. We
will also
take into account the six following coarse--grained histories:
\begin{equation}
	\hisxxpp\quad =\quad \hisxxpyymzzp\vee\hisxxpyypzzm, \label{eq17}
\end{equation}
\begin{equation}
	\hisxxmm\quad =\quad \hisxxmyypzzp\vee\hisxxmyymzzm, \label{eq18}
\end{equation}
coinciding with those appearing in Family A, \begin{equation}
	\hisyypp\quad =\quad \hisxxmyypzzp\vee\hisxxpyypzzm, \label{eq19}
\end{equation}
\begin{equation}
	\hisyymm\quad =\quad \hisxxpyymzzp\vee\hisxxmyymzzm, \label{eq20}
\end{equation}
coinciding with those appearing in Family B, \begin{equation}
	\hiszzpp\quad =\quad \hisxxpyymzzp\vee\hisxxmyypzzp, \label{eq21}
\end{equation}
\begin{equation}
	\hiszzmm\quad =\quad \hisxxpyypzzm\vee\hisxxmyymzzm, \label{eq22}
\end{equation}
which coincide with those appearing in family C. Note that the above
relations imply:
\begin{equation}
\hiszzpp\quad =\quad\{\hisxxpp\wedge\hisyymm\}\vee
	\{\hisxxmm\wedge\hisyypp\},
 \label{eq23} \end{equation}
\begin{equation}
\hiszzmm\quad =\quad\{\hisxxpp\wedge\hisyypp\}\vee
	\{\hisxxmm\wedge\hisyymm\},
 \label{eq24} \end{equation}
and that, obviously, the corresponding relations hold between their images
under
the homomorphism.

Given these premises we can prove our theorem. Let us consider the history
 $\hisuxpdxp$ belonging to family A, and let us assume that the spin component
 of particle 1 along the $x$ axis possesses the value +1 and that the same
hold
for the spin of particle 2. This means that
the history $\hisuxpdxp$ is true: $h\{\hisuxpdxp\} = 1$, and that the three
histories $\hisuxmdxp$, $\hisuxpdxm$ and $\hisuxmdxm$ are
 false\footnote{Of course, any other choice of the eigenvalues of the two
spin operators
will lead to the same contradiction as the one we will derive.}:
 $h\{\hisuxmdxp\} = 0$, $h\{\hisuxpdxm\} = 0$ and $h\{\hisuxmdxm\} = 0$. The
truth values of the histories $\hisxxpp$ e $\hisxxmm$ are then uniquely
determined by
the properties of the homomorphism $h$:

\begin{eqnarray*}
	h\{\hisxxpp\} & = & h\{\hisuxpdxp\vee\hisuxmdxm\} = \\ 	& =
 & h\{\hisuxpdxp\}\vee h\{\hisuxmdxm\} = \\ 	& = & 1 \vee 0 = 1,
\end{eqnarray*}
\begin{eqnarray*}
	h\{\hisxxmm\} & = & h\{\hisuxpdxm\vee\hisuxmdxp\} = \\
	& = & h\{\hisuxpdxm\}\vee h\{\hisuxmdxp\} = \\ 	& = & 0 \vee 0 = 0.
\end{eqnarray*}
The conclusion of our analysis can be summarized in the following table:

\begin{center}
\begin{tabular}{|c|c|c|c|c|c|} \hline
	$\hisuxpdxp$ & $\hisuxpdxm$ & $\hisuxmdxp$ & $\hisuxmdxm$ &
	$\hisxxpp$ & $\hisxxmm$ \\ \hline
	1 & & & & 1 & \\ \hline
	& 0 & 0 & 0 & & 0 \\ \hline
\end{tabular}
\end{center}
Now we take into account Family B and, without paying any attention to the
conclusions
 we have reached arguing within the previous family, we suppose that
particle 1 has its spin pointing along the positive direction of the axis
$y$, while
particle 2 has its spin pointing in the negative direction of the same
axis. We get then
another table:

\begin{center}
\begin{tabular}{|c|c|c|c|c|c|} \hline
	$\hisuypdyp$ & $\hisuypdym$ & $\hisuymdyp$ & $\hisuymdym$ & 
	$\hisyypp$ &
 $\hisyymm$ \\ \hline
	& 1 & & & & 1 \\ \hline
	0 & & 0 & 0 & 0 & \\ \hline
\end{tabular}
\end{center}
Analogous procedures can be applied to Family D: \begin{center}
	\begin{tabular}{|c|c|c|c|c|c|} \hline
	$\hisuxpdyp$ & $\hisuxpdym$ & $\hisuxmdyp$ & $\hisuxmdym$ &
	$\hisxypp$ & $\hisxymm$
 \\ \hline
	& 1 & & & & 1 \\ \hline
	0 & & 0 & 0 & 0 & \\ \hline
\end{tabular}
\end{center}
and to Family E:

\begin{center}
	\begin{tabular}{|c|c|c|c|c|c|} \hline
	$\hisuypdxp$ & $\hisuypdxm$ & $\hisuymdxp$ & $\hisuymdxm$ &
	$\hisyxpp$ & $\hisyxmm$
 \\ \hline
	1 & & & & 1 & \\ \hline
	& 0 & 0 & 0 & & 0 \\ \hline
\end{tabular}
\end{center}

We come now to dicuss Family C. As already remarked it contains two histories
$\hisxypp$ and $\hisxymm$ which coincide with two histories belonging to
Family D;
according to assumption c) of Section 2 they must have the same truth
values. An analogous
argument holds for the histories $\hisyxpp$ e $\hisyxmm$. By considering
the truth values
of all these histories and taking into account relations (\ref{eq11}) and
(\ref{eq12}),
we can deduce the truth values of the histories $\hiszzpp$ and $\hiszzmm$:
\begin{eqnarray}
	h\{\hiszzpp\} & = &
	h\{[\hisxypp\wedge\hisyxpp]\vee[\hisxymm\wedge\hisyxmm]\}
	\nonumber \\
	& = & [h\{\hisxypp\}\wedge h\{\hisyxpp\}]\vee 	\nonumber \\
	& & \vee[h\{\hisxymm\}\wedge h\{\hisyxmm\}] \label{eq25} \\ 	& = &
[0 \wedge 1]\vee[1 \wedge 0] = 0 \vee 0 = 0, \nonumber \end{eqnarray}
\begin{eqnarray}
	h\{\hiszzmm\} & = &
	h\{[\hisxymm\wedge\hisyxpp]\vee[\hisxypp\wedge\hisyxmm]\}
	\nonumber \\ & =
 & [h\{\hisxymm\}\wedge h\{\hisyxpp\}]\vee \nonumber \\ 	& &
 \vee[h\{\hisxypp\}\wedge h\{\hisyxmm\}] \label{eq26} \\ 	& = & [1
\wedge 1]\vee[0
 \wedge 0] = 1
\vee 0 = 1. \nonumber
\end{eqnarray}

As one should have expected the two truth values are opposite, since the
two considered
 histories are mutually exclusive and exhaustive. In this way we have identified
the truth table for the histories of Family C:

\begin{center}
\begin{tabular}{|c|c|c|c||c||c||} \hline $\hisxypp$ 	& $\hisxymm$ &
$\hisyxpp$ &
 $\hisyxmm$ & $\hiszzpp$ & 	$\hiszzmm$ \\ \hline
	& 1 & 1 & & & 1 \\ \hline
	0 & & & 0 & 0 & \\ \hline
\end{tabular}
\end{center}
The last step consists in performing a similar analysis for Family F. As
already remarked
its two histories $\hisxxpp$ e $\hisxxmm$ coincide with histories belonging
to Family A and,
 according to assumption c), must have the same truth values, 1 and 0,
respectively. The
same holds for the histories $\hisyypp$ and $\hisyymm$, which coincide with
two histories
 belonging to Family B. Just as in the previous case,
taking into account the relations (\ref{eq23}) and (\ref{eq24}), we can
then evaluate the
 truth values of the two histories
$\hiszzpp$ and $\hiszzmm$:
\begin{eqnarray}
	h\{\hiszzpp\} & = &
	h\{[\hisxxpp\wedge\hisyymm]\vee[\hisxxmm\wedge\hisyypp]\} \nonumber
\\ 	& = &
 [h\{\hisxxpp\}\wedge h\{\hisyymm\}]\vee \nonumber \\ 	& &
\vee[h\{\hisxxmm\}\wedge h\{\hisyypp\}] \label{eq27} \\ 	& = & [1
\wedge 1]\vee[0 \wedge 0] = 1 \vee 0 = 1, \nonumber
\end{eqnarray}
\begin{eqnarray}
	h\{\hiszzmm\} & = &
	h\{[\hisxxpp\wedge\hisyypp]\vee[\hisxxmm\wedge\hisyymm]\} \nonumber
\\ 	& = &
[h\{\hisxxpp\}\wedge h\{\hisyypp\}]\vee \nonumber \\
	& & \vee[h\{\hisxxmm\}\wedge h\{\hisyymm\}] \label{eq28} \\ 	& = &
 [1 \wedge 0]\vee[0 \wedge 1] = 0 \vee 0 = 0. \nonumber
\end{eqnarray}
We can then exhibit the truth table for the histories of Family F:

\begin{center}
\begin{tabular}{|c|c|c|c||c||c||} \hline 	$\hisxxpp$ & $\hisxxmm$ &
$\hisyypp$ &
 $\hisyymm$ & $\hiszzpp$ & 	$\hiszzmm$ \\ \hline
	1 & & & 1 & 1 & \\ \hline
	& 0 & 0 & & & 0 \\
\hline
\end{tabular}
\end{center}
Comparing the two last truth tables one sees that the Families C and F
attribute opposite
truth values to the two histories $\hiszzpp$ e $\hiszzmm$: if one limits
his considerations
to Family C, then we can claim with certainity that both particles have
their spin antiparallel along the $z$ axis, on the contrary, if we take
into consideration
Family F, then we must conclude that the two particles have their spin
parallel with respect to the same axis.

To avoid being misunderstood we stress once more that we have used only the
four assumptions
listed in the previous section to derive the above contradiction; in
particular we have never
 made statements involving {\it different} histories belonging to
incompatible decoherent families.

\section{An answer to Griffiths' objections.}

R. Griffiths\footnote{Private communication.} has repeatedly expressed (in
private correspondence) his disappointment with this paper, claiming
that in developing our argument we violate one of the fundamental
rules of the DH approach, the one we have already mentioned and
which we shell refer to as   {\it the
single family rule} in what follows. According to such a rule {\it any
reasoning must employ a single family of decoherent histories}. Since in
the crucial example of Section 3 we resort to the consideration of six
incompatible decoherent families and we combine, in a way or
another, the conclusions drawn within the six families to derive a
contradiction, we violate the single family rule; consequently, in
Griffiths' opinion, our line of thought is wrong. 

Griffiths has also repeatedly called our attention on an example which
has been exhaustively discussed in the literature. In his opinion such an
argument is strictly similar to ours (\cite{gh1} and references therein) but
it is much simpler and it should allow to better understand the essential
role of the single family rule as well as its
implications\footnote{Goldstein
\cite{gold} has considered a similar but slightly more elaborated
example. The following remarks hold both for Griffiths' as well as for
Goldstein's examples.}. The rule and its implications are, in Griffiths'
opinion, accepted both by the supporters as well as by the opponents of
the DH approach.

In brief the example of ref. \cite{gh1} which we present here in a slightly
modified version used by Griffiths in our correspondence 
goes as follows: one considers three orthogonal states 
$|A\rangle, |B\rangle$ and
$|C\rangle$, and the two states:
\begin{eqnarray}
	|\varphi\rangle & = & \frac{1}{\sqrt{3}}\left[ |A\rangle + |B\rangle +
	|C\rangle \right], \\
	|\psi\rangle & = & \frac{1}{\sqrt{3}}\left[ |A\rangle + |B\rangle -
	|C\rangle \right]. 
\end{eqnarray}
Let  $t_{0} < \tu < \td$ be three time instants  and suppose that the dynamics 
is
trivial, i.e. $H=0$. One then considers the following three histories: 
\begin{eqnarray}
	\hisu & = & \left\{(|\varphi\rangle\langle\varphi|, t_{0}), (1, \tu),
	(|\psi\rangle\langle\psi|, \td)\right\}, \\
	\hisd & = & \left\{(|\varphi\rangle\langle\varphi|, t_{0}),
	(|A\rangle\langle A|, \tu), (1, \td)\right\}, \\
	\hist & = & \left\{(|\varphi\rangle\langle\varphi|, t_{0}),
	(|B\rangle\langle B|, \tu), (1, \td)\right\}.
\end{eqnarray}
It is then possible to prove that there is a decoherent family
${\mathcal A}$
such that the two histories  $\hisu$ and $\hisd$ belong to it and,
moreover, within it:
\begin{equation} \label{e37}
\frac{p(\hisu\wedge\hisd)}{p(\hisu)} = 1
\qquad \makebox{so that} \qquad
\hisu \Rightarrow \hisd,
\end{equation}
$p( )$ being the probability distribution characterizing the histories
of the family
${\mathcal A}$.

There is also a second family of decoherent
histories
${\mathcal B}$ such that both  $\hisu$ and $\hist$ belong to it and,
within it, one has:
\begin{equation} \label{e38}
\frac{p(\hisu\wedge\hist)}{p(\hisu)} = 1
\qquad \makebox{so that} \qquad 
\hisu \Rightarrow \hist.
\end{equation}
Obviously, in the above equation $p( )$ is the probability distribution
characterizing the histories of
${\mathcal B}$. 

Finally, there is a third decoherent family 
${\mathcal C}$ which accomodates $\hisd$ and $\hist$. Obviously, such
histories are  mutually exclusive since the states  $|A\rangle$ and
$|B\rangle$ are orthogonal: this implies that they cannot correspond
simoultaneously to physical properties of the system under
consideration.

Let us now suppose that the history
$\hisu$ is true. According to (\ref{e37}) we can conclude that also
$\hisd$ is true, and, according to  (\ref{e38}) that  $\hist$
must also be true. This, however, cannot happen since the two considered
histories are mutually exclusive. 

Why is this argument not correct? As
remarked by the DH supporters and in particular by Griffiths, the argument
violates the single family rule since the conclusion requires the
consideration of three different decoherent families  ${\mathcal A},
{\mathcal B}$ and
${\mathcal C}$, which are incompatible with each other. The line of
reasoning leading to the contradiction is forbidden by the rules of the
DH 
approach.

Even though we believe, with  d'Espagnat
\cite{de1, de2}, that the single family rule cannot be considered so
natural and free from puzzling aspects as the  DH supporters seem
to believe, we are perefectly aware that, at the purely formal
level,  Griffiths' criticisms concerning the just discussed example are
legitimate. Does this impliy that 
the same conclusion holds also for 
our example of Section 3? We belive that this is not the case: 
\begin{itemize}
\item Since Griffiths has repeatedly stated (in his papers and in his
correspondence with us) that decoherent histories refer to objective
properties of physical systems, that they are the {\it beables} of the
DH approach, then he  
{\it must} accept that  they have  precise truth values. The very existence
of a  truth--functional for the histories of a decoherent family amounts
simply to the formal expression that such histories speak of objective
properties of physical systems:
\begin{center}
\begin{tabular}{|c|}
	\hline
	\\
	Decoherent histories refer to {\it objective} properties of physical
	systems. \\
	$\Downarrow$ \\
	They are given a precise truth value: 0 or 1. \\
	\\
	\hline
\end{tabular}
\end{center}
Moreover if we want that within a decoherent family the usual
classical rules can be used (once more Griffiths himself, as well as 
Omn\`es, have repeatedly stressed the necessity of this feature), then we
must accept that such a  truth--functional be a homomorphism:
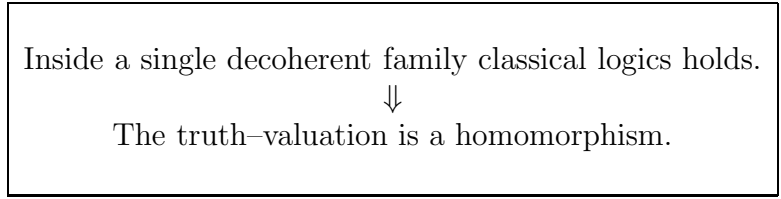
\begin{center}
\begin{tabular}{|c|}
	\hline
	\\
	Inside a single decoherent family classical logics holds. \\
	$\Downarrow$ \\
	The truth--valuation is a homomorphism. \\
	\\
	\hline
\end{tabular}
\end{center}
All this has been described and discussed in details in
Section 2. Thus, it seem useless to us that Griffiths insists, in his
correspondence, that within his theory decoherent histories speak of
objective properties of physical systems, but that no homomorphism of the
kind we have just envisaged exists. Such an attitude is
contradictory: either decoherent histories make reference to objective
properties, but then one must unavoidably accept the existence of a
truth valuation (which must be an homomorphism if classical rules
must hold) for them, or one denies the very possibility of considering a
truth valuation, but then the histories loose any objective physical
meaning. He can make the choice he prefers.

\item We stress that the existence of a truth value for the
histories of a decoherent family (if one considers them as referring
to objectively possessed properties) preceeds logically (actually
ontologically) the assignement of a probability distribution to them.
Actually, the probability distribution makes reference to our 
knowledge about the physical system which has a fundamentally
contingent character, in the precise sense that it depends on the
information we have at a considered time instant. This cannot change
the fact that the physical system has objective properties (accordingly, the
histories are true if they account for such properties and are false if they
do not) which are completely independent from the probabilistic
(epistemic) informations we have about the system. Since the properties
--- and not the probability --- correspond (in a realistic position as the
one advocated by Griffiths) to physical reality, they must play a primary
role within the theory and should be the objects of  interest for the
scientist.

\item Consideration of the  Kochen and Specker theorem raises the problem of
whether the mathematical formalism of quantum mechanics allows a consistent
assignement of truth values to the projection operators of the Hilbert space,
i.e. whether they can be considered as representing  {\it objective}
properties of physical systems,  {\it indipendenently} of the probabilities
one attaches to such projection operators. Accordingly, the theorem deals
with  the algebraic properties of projection operators (with specific
reference to their non commutative nature) but it never takes into
consideration the probabilities of the formalism. According to
our previous analysis it should be clear why 
Kochen and Specker have chosen this line of approach: if a theory pretends
to speak of properties objectively possessed by physical systems, such
properties, just because they are  {\it objective} cannot depend on our
probabilistic knowledge which, in general, has a   {\it
subjective} character\footnote{Obviously the independence of
objective properties from probability distributions cannot be
complete, since, e.g., in the case in which the theory attaches
probability 1 to a property, then it must be possessed.}.  

Our example of Section 3 can be considered as a simple transcription of the 
Kochen and Specker theorem (in the version of Peres and Mermin) in the
language of decoherent histories. Our aim is to prove that also within such
a theoretical scheme one cannot attribute too many  {\it
objective} properties to physical systems, just as in standard quantum
mechanics. From this point of view, it has to be stressed that in our
reasoning we  {\it never} make reference to the probability distributions
which can be attached to the histories of the six considered families.
Viceversa, in Griffiths' example   {\it the whole argument is based in a
fundamental way on the consideration of the probability distribution},
as it is evident from equations 
(\ref{e37}) and (\ref{e38}). This is an important difference between the
two arguments which, in our opinion, Griffiths has not been able to
grasp, in spite of the fact that it is evident that one cannot derive a  
Kochen and Specker--like contradiction by resorting to the 
example he takes into account. This is not a purely formal difference; it has
precise implications, as we are going to discuss.

\item The ``single family rule'',  as we have already stated, is the 
{\it ``fundamental principle of quantum resoning''} \cite{gri2}. It states
that:  
\begin{quote}
{\small Any reasoning must emply a single family of decoherent histories.}
\end{quote}
Griffiths himself has made very clear what he means by the expression 
{\it ``quantum  resoning''}
\cite{gri2}:
\begin{quote}
{\small The type of quantum resoning we shall focus ... is that in which one
starts with some information about a system, known or assumed to be true,
and from these {\it initial data} tries to reach valid {\it conclusions}
which will be true if the initial data are correct.}
\end{quote}
As one can easily grasp from his last papers 
\cite{gri2},
when he uses the expression {\it ``quantum resoning''} he has actually in
mind a {\it reasoning of exclusively probabilistic nature} (due to the fact
that we can have only a probabilistic knowledge about physical systems),
which allows to manipulate probabilities to derive new informations. From
this point of view the   single
family rule might appear as a reasonable request: actually, since the
 ``quantum reasoning'' has a fundamentally probabilistic nature, and since
the probabilities depend from the considered family (just as in classical
mechanics they depend on the graining one choses, and different grainings
correspond to different probability assignements) it seems natural that 
``quantum reasoning'' (in the above sense) depends from the decoherent
family one considers.

However, as repeatedly stressed, if one of the basic assumptions of
the theory is that histories refer to objective properties of
physical systems, then  {\it beside a fundamentally probabilistic  ``quantum
resoning'' based on our knowledge about the system, there must be a
second type of reasoning based on the properties objectively possessed
by physical systems, independently from our precise knowledge  of them}.
This fact, we stress it once more, is imposed by the very statement that the
decoherent histories speak of objective properties of physical systems. This
is precisely the reasoning at the basis of the   Kochen and Specker
theorem (even though their interest was directed to hidden variable
theories). And this is precisely the reasoning of our Section 3. The two
ways of reasoning are fundamentally different and there is no reason for
which the    single family rule, which might hold for the first line of
thought, must obviously hold also for the second. On the contrary, it
cannot hold for the second one since assuming such a rule within this
perspective amounts simply to assume that the physical properties are not
objective, being related to the family which one chooses.
\end{itemize}

The argument of Section 3 is different, not only for its formal aspects, but
for its very essence from those which have been considered in the
literature, and, in particular, from the one which is repeatedly mentioned
by Griffiths. For these reasons, our argument cannot simply be dismissed by
invoking acritically the single family rule, as it has happened up to now.

\section{Conclusions.}

The conclusion of our investigation should be obvious: if one wants to
entertain the
 Decoherent Histories point of view, he must give up at least one of the
previous assumptions.
Let us discuss a little bit more what happens if we relax one of them. \\

$\bullet$ If one gives up the request that any decoherent familly be endowed
with a
boolean structure, then he is giving up the possibility of using classical
reasoning within
such a family, loosing in this way the nicest feature of the theory and the
very reason to
consider it. Since, as stated before, nobody seems to contemplate this
possibility, we do not
 discuss it any further. \\

$\bullet$ One could give up the second assumption, stating that not every
decoherent history
 has a truth value. This is, in our opinion, a very dangerous move: in fact,
giving a truth value to a decoherent history is not simply a formal act,
but it means that we
 are establishing a precise correspondence between such a history and some
objective physical
 properties. If we deny any truth value to the history, then we deny such
correspondence, and
the history becomes just an empty statement devoid of any physical meaning.
In Classical
 Statistical Mechanics, all events in phase space are
given a truth value, because they all correspond to particular physical
properties, even if
one in general knows only their probability distributions. In Standard
Quantum Mechanics, on
the other hand, no truth--value assignment exists in general, and in fact
the quantum
 projection operators do not correspond in general to any physical property
possessed by
systems for the simple reason that quantum systems do not have actual but
only potential
properties before a measurement process is performed. So if we assume that
some histories have
 no truth value, then we must accept that they are meaningless from the
physical point of
 view. Of course, this is not a problem, but then the theory has to tell us
which histories
have a truth value, and which do not, i.e. which correspond to physical
properties (and then
 have a precise ontological status) and which do not (and, as such, are
only empty statements
 devoid of any ontological meaning): without any such prescription, the
theory would be
incomplete.

Omn\`es \cite{omn1a, omn2}, for example, has tried to give a precise answer
to the previous
 question: specifically,
he has proposed a criterion for truth which is independent from the
families, and which also
 eliminates the problem of the existence of families describing senseless
properties for
classical macroscopic objects. Unfortunately, Dowker and Kent \cite{dk}
have shown that his
 proposal is not tenable. \\

$\bullet$ Assumption c) seems to us impossible to give up\footnote{Of
course, if one accepts
that some histories have no truth value (i.e. he gives up assumption b)),
then assumption c)
 becomes meaningless for those histories.}. In fact, let us recall
the argument concerning the impossibility of considering, within hidden
variable theories,
 the values of the observables of the table at the beginning of Section 3
as uniquely
 determined by the hidden variables (or equivalently, as objectively
possessed). There,
 we have mentioned that the only consistent way out
from this embarrassing situation derives from accepting that
the truth values of statements concerning the predictions of the theory
about the outcomes of
 measurements depend from the whole context. In particular, different truth
values are
 necessarily associated
to different and incompatible measurement procedures, i.e., to different
physical
situations. In the case of
the Decoherent Histories the situation is radically different. In fact,
they do not
speak of measurement outcomes but of properties possessed independently of
any procedure
 to test them. Therefore, within such a conceptual framework to make the
truth value of a
 precise history dependent from
the family to which {\it it is considered} to belong seems to us logically
unacceptable:
 it would be better to keep the Copenhagen interpretation. \\

$\bullet$ If we decide to give up assumption d), then we recognize that the
theory as
it stands is not complete, because the decoherence condition by itself does
not select
the proper families to be used for describing physical systems, and we have
to find new
criteria in order to complete the theory. This fact
does not mark by itself the definitive failure of the program: it simply
points out that
 the theory needs to be enriched by new assumptions apt to identify the
family, or the
families, which are physically significant. This, however, is not an easy
task, and our
 example throws a precise and disquieting light on
the difficulties one will meet in trying to consistently implement such ideas.
In fact we can raise the question:
which one (or ones) of the six families summarized in the table at the
beginning
of Section 3 should be discarded? Which criterion could one use making some
of these families
acceptable and forbidding the consideration of the remaining ones, given
the fact that they
 have a quite similar conceptual status and they speak of analogous
properties of our
 system? \\

To conclude, our analysis shows that the DH approach, as it stands
(whatever interpretation
 one decides to subscribe), is either incomplete or does not meet the
requirements for
 a ``realistic'' description of the physical world, the very reason for
which it has been
 proposed.

\section*{Acknowledgments}

We acknowledge useful discussions with A. Kent and R. Griffiths.

\end{document}